\documentclass[12pt]{amsart}
\usepackage{amsmath}

\newtheorem{thm}{Theorem}[section]
\newtheorem{lem}[thm]{Lemma}%[section]
\newtheorem{prop}[thm]{Proposition}%[section]
\newtheorem{cor}[thm]{Corollary}%[section]
\theoremstyle{definition}

\theoremstyle{remark}
\newtheorem{remark}{Remark}[section] % \renewcommand{\theremark}{}

\theoremstyle{plain}

\newcommand{\C}{{\mathbf C}}

\newcommand{\HH}{\mathbf H}

\newcommand{\norm}[1]{\lVert #1\rVert}

\newcommand{\OPN}{\operatorname{Op}_N}

\newcommand{\FN}{\mathcal F_N}
\newcommand{\HN}{\mathcal H_N}
\newcommand{\UN}{\mathcal U_N}
\newcommand{\UaN}{\mathcal U_{a,N}}

\newcommand{\VaN}{\mathcal V_{a,N}}
\newcommand{\VoN}{\mathcal V_{0,N}}
\newcommand{\TN}{T_N}

\newcommand{\Fh}{\mathcal F_h} % h-Fourier transform

\numberwithin{equation}{section}

\def\HH{{\mathbb H}}

\def\RR{{\mathbb R}}
\def\TT{{\mathbb T}}
\def\ZZ{{\mathbb Z}}

\def\e{\mathrm{e}}
\def\i{\mathrm{i}}

\def\id{\operatorname{id}}

\def\C{\operatorname{C{}}}

\def\L{\operatorname{L{}}}

\def\Op{\operatorname{Op}}
\def\sOp{\widetilde{\operatorname{Op}}}

\def\PSL{\operatorname{PSL}}

\begin{document}

\title
{Quantum unique ergodicity for parabolic maps}
\author{Jens Marklof and Ze\'ev Rudnick}
\address{Institut des Hautes \'Etudes Scientifique, 35 route de Chartres,
91440 Bures-sur-Yvette, France {\em and}
Laboratoire de Physique Th\'eorique et Mod\`eles Statistiques, 
bat.~100, Universit\'e Paris-Sud, 91405 Orsay Cedex, France;
{\em Permanent address:} School of Mathematics, University of Bristol,
Bristol BS8 1TW, U.K. ({\tt J.Marklof@bris.ac.uk})}
\address{Raymond and Beverly Sackler School of Mathematical Sciences,
Tel Aviv University, Tel Aviv 69978, Israel
({\tt rudnick@math.tau.ac.il})}
\date{4 January 1999, revised 9 February 1999}
\thanks{JM is supported by the 
European Post-Doctoral Institute 
for the Mathematical Sciences and the European Commission 
(TMR Marie Curie Grant). Visits to Tel Aviv University have been
supported by the Hermann Minkowski Center for Geometry.}
\thanks{ZR is supported  in part by grants from 
the Israel Science Foundation  and an Alon fellowship.} 

\begin{abstract}
We study the ergodic properties of quantized  ergodic 
maps of the torus.  
It is known that these satisfy {\em quantum ergodicity}: For {\em
almost} all eigenstates, the expectation values of quantum
observables converge to the classical phase-space average with respect
to Liouville measure of the corresponding classical observable.

The possible existence of any exceptional subsequences of eigenstates
is an important issue, which until now was unresolved in any example. 
The absence of exceptional subsequences is referred to as 
quantum {\em unique} ergodicity (QUE).  
We present the first examples of maps which satisfy QUE: 
Irrational skew translations of the two-torus,
the parabolic analogues of Arnold's cat maps. 
These maps are classically uniquely ergodic and not mixing.
A crucial step is to find a quantization recipe which
respects the quantum-classical correspondence principle. 

In addition to proving QUE for these maps, we also give results on the
{\em rate of convergence} to the phase-space average. We give upper bounds
which we show are optimal. We construct special examples of these maps
for which the rate of convergence is arbitrarily slow. 
\end{abstract}

\maketitle

%\tableofcontents

%\input{intro}    
\section{Introduction}
% alternative 

One of the central problems of ``Quantum Chaos'' is the question of
the asymptotic behaviour of eigenmodes of classically chaotic systems
in the semiclassical limit. In particular, one wants to find their
limiting ``mass distribution'' in a suitable sense. 

Consider for instance the geodesic flow on a compact Riemannian manifold $M$ 
(or rather, on its co-tangent bundle), 
whose quantum Hamiltonian is, in suitable units, represented by $-\Delta$,
the positive Laplacian on $M$.
Let  $\psi_j$ be a sequence of normalized 
eigenfunctions: $\Delta \psi_j +\lambda_j \psi_j=0$, 
$\int_M |\psi_j|^2 = 1$.  
A suitable quantity for measuring the concentration properties of the
eigenmodes $\psi_j$, 
in both the position and momentum representations, is the
distribution on the unit co-tangent bundle $S^*M$ given by 
\begin{equation}\label{dphih}
f\in \C^\infty(S^* M) \mapsto 
\langle \Op(f) \psi_j,\psi_j \rangle .
\end{equation}
Here $\Op(f)$ is a zero-order pseudo-differential operator with
principal  symbol $f\in \C^\infty(S^* M)$ and $\Op$ is some choice of
quantization from symbols to pseudo-differential operators. 
The operator $\Op(f)$ is a quantization of the classical observable
$f$, and  
$\langle \Op(f) \psi_j,\psi_j \rangle$ are the expectation values for
the operator in the state $\psi_j$. 

A celebrated result in this direction  is 
``Schnirelman's theorem''\footnote{announced in \cite{Schnirelman74}
with full proofs given by Zelditch \cite{Zelditch87} for hyperbolic
surfaces and Colin de Verdiere \cite{CdV85} in general, see
also \cite{Helffer87}.} 
which says that if the flow is {\em ergodic}  
then these expectations converge to the phase-space average of the
classical  observable $f$, for all but possibly a zero-density
subsequence of eigenfunctions.    
This phenomenon is commonly referred to as 
{\em quantum ergodicity}\footnote{There are other notions
of ergodicity in quantum mechanics, such as von Neumann's 
\cite{vonNeumann29,Klimek97}, which are not related to the one used here.}.    
There are no examples where it is known if there are any 
exceptional subsequences. 
The case where there are none is referred to as 
{\em quantum unique ergodicity} (QUE) \cite{Rudnick94,Luo95,Jakobson94}.  

In this paper, we consider a compact model of the above situation, 
where the dynamics, instead of taking place in the co-tangent bundle,  
occurs in a {\em compact} symplectic manifold, namely the 2-torus 
$\TT^2$. The (classical) evolution is then given by iterating 
a symplectic map of the torus. 

In order to quantize such a map, one looks
for a Hilbert space  of state-vectors of the system, which are
required to be periodic in both position and momentum representations.
This constrains Planck's constant $h$ to be an inverse integer: $h=1/N$, 
and then the state space $\HN$ is finite dimensional, 
of dimension precisely $N$.  
The semiclassical limit is now $N\to \infty$. 
By means of an analogue of Weyl quantization, one defines quantum 
observables $\OPN(f)$ corresponding to smooth classical observables
$f\in \C^\infty(\TT^2)$. 

Given a symplectic map $A$ of $\TT^2$, 
the quantum evolution is given by specifying a unitary operator $\UN$
on the state space $\HN$, which satisfies a version of the 
``correspondence principle''  (Egorov's theorem): 
\begin{equation}\label{integorov}
\norm{\UN^{-1}\, \OPN(f)\, \UN - \OPN(f\circ A)} 
\stackrel{N\rightarrow\infty} \longrightarrow  0,\quad
\forall f\in \C^\infty(\TT^2) ,
\end{equation}
where $f\circ A(\begin{smallmatrix}p\\q\end{smallmatrix}) =
f(A(\begin{smallmatrix}p\\q\end{smallmatrix}))$,  
that is one requires that in the semiclassical limit, quantum
evolution becomes classical evolution. 
The analogue of eigenmodes are then the eigenfunctions of the propagator 
$\UN$. 

The main focus in the literature has so far been on 
{\em hyperbolic} transformations of the torus, the so-called
cat maps \cite{Hannay80,Keating91,Degli93,Degli95}, 
to which the proof of Schnirelman's theorem \cite{Schnirelman74,
Zelditch87, CdV85} can be adapted to prove quantum ergodicity, but not QUE
\cite{Bouzouina96,Zelditch97}. Assuming the Generalized Riemann
Hypothesis, Degli~Esposti, Graffi and Isola
\cite{Degli95} found an explicit infinite (though sparse) subsequence
of values of  $N$, for which they show that the expectation values for
{\em all} eigenfunctions $\psi\in\HN$  converge to the phase space average.

Here, we will study a {\em parabolic} map of the torus (also called a skew
translation), which is specified by choosing 
a real number $\alpha$, and then defining\footnote{For a technical
reason we shift $q$ by $2p$ rather than $p$.}    
$$
A_\alpha: \begin{pmatrix} p\\ q\end{pmatrix} \mapsto 
\begin{pmatrix}p+\alpha\\ q+2p \end{pmatrix} \bmod 1 .
$$

When $\alpha=0$ the motion is clearly integrable as $p$ is a constant
of the motion. For {\em rational} values of  $\alpha$, the map is
``pseudo-integrable'' in that the dynamics of the map on an orbit 
can be identified with an interval exchange transformation. 
For $\alpha$ irrational, the map is {\em ergodic} and in fact it was 
found by Furstenberg \cite{Furstenberg61} to be {\em uniquely} ergodic. 
These maps possess no further ``chaotic'' properties; 
for instance they are not mixing.

We propose a quantization procedure that at each value of $N$,
replaces $\alpha$ by 
a rational approximant $a/N$, and then construct a unitary propagator
$\UaN$ on $\HN$ which satisfies an exact version of Egorov's Theorem: 
$$
\UaN^{-1}\, \OPN(f)\, \UaN = \OPN(f\circ A_{a/N}) .
$$
Then taking any sequence $a/N\to \alpha$ we show that 
\eqref{integorov} holds.
This gives us a quantization of the map $A$. There are other recipes
in the literature \cite{DeBievre96,Bouzouina96}; 
however, they do not satisfy \eqref{integorov}. 

Once we have the analogue of Egorov's theorem \eqref{integorov} and
have set up the necessary tools from 
pseudo-differential calculus on $\TT^2$,  we show:
\begin{thm}[QUE for parabolic maps]\label{abs thm} 
Suppose $\alpha$ is irrational, $f\in \C^\infty(\TT^2)$ a smooth 
observable, and $a/N\to \alpha$ a sequence of rational approximants. 
Then for any normalized eigenfunctions $\psi\in \HN$ of the propagator
$\UaN$, we have 
$$
\langle \OPN(f) \psi,\psi \rangle \to \int_{\TT^2}f(p,q)\, dp\, dq ,
\qquad N\to\infty .
$$
\end{thm}
That is the parabolic map $A_\alpha$ satisfies  
{\em quantum unique ergodicity}. This is the first known example of QUE.

The remainder of our paper concerns the {\em rate} of convergence. 
We take approximants so that $|\alpha-a/N| < 1/N$. 
Suppose that $\alpha$ is badly approximable  (in the sense that 
$|\alpha -p/q|\gg_\epsilon q^{-2-\epsilon}$ for all $\epsilon>0$). 
We then show (Corollary~\ref{cor badly app}) that for any normalized 
eigenfunction $\psi$ of the propagator $\UaN$ we have
$$ 
|\langle \OPN(f)\psi,\psi \rangle -\int_{\TT^2}f(p,q)\, dp\, dq |
\ll_{f,\epsilon} N^{-1/4+\epsilon},\quad \forall \epsilon>0 .
$$
The reason why the rate is $N^{-1/4+\epsilon}$ and not, 
as one might have guessed $N^{-1/2+\epsilon}$, are degeneracies 
in the spectrum,
which occur whenever $a$ and $N$ are not co-prime. 
We can, however, always construct an explicit basis of eigenfunctions
$\psi_j$ ($j=1,\ldots,N$), for which
$$
|\langle \OPN(f)\psi_j,\psi_j \rangle -\int_{\TT^2}f(p,q)\, dp\, dq |
\ll_{f} N^{-1/2} ,
$$
see Section \ref{explicit sec}.
In the absence of degeneracies
($a$ and $N$ co-prime), we thus indeed obtain a rate of $N^{-1/2}$
(cf.~also Theorem \ref{thm upper}).

As for lower bounds on the rate, 
we show (Theorem~\ref{thm lower all}) 
that for the observable $f(p,q) = \e^{2\pi\i \cdot 2p}$, for
all irrationals there is a sequence of values of $N$ and normalized
eigenfunctions $\psi\in \HN$ for which 
$$
|\langle \OPN(f)\psi,\psi \rangle -\int_{\TT^2}f(p,q)\, dp\, dq | 
\gg \frac 1{N^{1/4}} .
$$
Thus for badly approximable $\alpha$, Corollary~\ref{cor badly app} 
is sharp. 
Moreover, unlike the situation with badly approximable $\alpha$, 
we can construct irrationals 
for which the rate of convergence in Theorem~\ref{abs thm}  is arbitrarily
slow, e.g. slower then $1/\log\log\log N$ (Theorem~\ref{thm lower}).

\newpage

\section{Quantum mechanics on $\TT^2$}

\subsection{Notation}
We write $e(x)=\e^{2\pi\i x}$ and $e_N(x)=\e^{\frac{2\pi\i}{N} x}$.
$\ZZ_N$ denotes the residue class ring $\ZZ/N\ZZ$.
$A\ll_{\epsilon} B$ and $A=O_{\epsilon}(B)$ both mean that there is 
a positive constant 
$c$ depending only on $\epsilon$, such that $|A|\leq c |B|$.

\subsection{The Hilbert space of states}
To recall the basics of quantum mechanics on the compact phase-space
$\TT^2$ \cite{Hannay80,Degli93,Degli95},   
we begin by describing the Hilbert space of states of such a system. 
The guiding rule is Heisenberg's uncertainty principle, which 
asserts that simultaneous 
measurements of momentum $p$ and position $q$ of a quantum particle
are ambiguous within Planck cells of volume $h$ (Planck's constant).
Hence if the phase space volume $V$ is finite, the dimension $N$
of the Hilbert space $\HN$ describing the state of the quantum particle
has to be finite as well, and is precisely given by $N=V/h$. 

In the case of the torus $\TT^2$, we take  state vectors to be
distributions on the line which are periodic in both momentum and
position representations: $\psi(q+1)=\psi(q)$,
$[\Fh\psi](p+1)=[\Fh\psi](p)$, where 
$[\Fh\psi](p) = h^{-1/2}\int\psi(q)\,e(-pq/h)\,dq$.  
The space of such distributions is finite
dimensional, of dimension precisely $N=1/h$, and consists of
periodic point-masses at the coordinates $q=Q/N$, $Q\in \ZZ$. 

We may then identify $\HN$ with the 
$N$-dimensional vector space $\L^2(\ZZ_N)$, with
the inner product $\langle\,\cdot\,,\,\cdot\,\rangle$ defined by
\begin{equation}
\langle \phi,\psi \rangle 
 = \frac1N \sum_{Q\bmod N} \phi(Q) \, \overline\psi(Q) ,
\end{equation}
This inner product induces a norm $\norm{\cdot}$ on the space of operators
on $\HN$, that is on the space of $N\times N$ matrices.

The Fourier transform $\FN$ may now be defined as
the unitary map
\begin{equation}
\widehat\psi(P) = [\FN \psi] (P) = N^{-\frac12} \sum_{Q\bmod N}
\psi(Q) \, e_N(-QP) ,
\end{equation}
its inverse $\FN^{-1}$ is then clearly given by
\begin{equation}
\psi(Q) = [\FN^{-1} \widehat\psi] (Q) = N^{-\frac12} \sum_{P\bmod N}
\widehat\psi(P) \, e_N(PQ).
\end{equation}

\subsection{Translation operators}
A central role will be played by the translation operators
$$
[t_1 \psi](Q) = \psi(Q+1)
$$
and
$$
[t_2 \psi](Q) = e_N(Q)\, \psi(Q),
$$
which may be viewed as the analogues of differentiation and
multiplication (respectively) operators in usual Fourier analysis on $\RR^n$. 
In fact in terms of the usual translation operators on the line 
$\hat q \psi(q)=q\psi(q)$ and 
$\hat p\psi(q)=\frac{h}{2\pi\i} \frac{d}{dq}\psi(q)$, they are given by    
$t_1=e(\hat p)$, $t_2=e(\hat q)$. 
Heisenberg's commutation relations read in this context
\begin{equation} \label{commrel}
t_1^a t_2^b = t_2^b t_1^a e_N(ab)  \qquad \forall a,b\in\ZZ.
\end{equation}

The Fourier conjugates of $t_1$ and $t_2$ are
$$
\FN t_1 \FN^{-1} = t_2
$$
and
$$
\FN t_2 \FN^{-1} = t_1^{-1} .
$$

\subsection{Observables}

For $n=(n_1,n_2)\in\ZZ^2$ put
$$
T_N(n) = e_N(\frac{n_1 n_2}{2}) t_2^{n_2} t_1^{n_1} .
$$
Then 
$$
T_N(m)\,T_N(n)=e_N(\frac{\omega(m,n)}{2}) \, T_N(m+n)
$$
with the symplectic form
$$
\omega(m,n)=m_1 n_2 - m_2 n_1 .
$$

For any smooth function $f\in\C^\infty(\TT^2)$ on our
phase space $\TT^2$, define a {\em quantum observable}
$$
\Op_N(f) = \sum_{n\in\ZZ^2} \widehat f(n) T_N(n)
$$
where $\widehat f(n)$ are the Fourier coefficients of $f$.
The observable $\Op_N(f)$ is also called the {\em Weyl
quantization of $f$}.

We have $\OPN(f)^* = \OPN(\bar f)$ and hence $\OPN(f)$ is self-adjoint
for real-valued $f$. 

The connection of these quantum observables
with the ``classical'' translations of the torus
$$
S_1^\alpha : \begin{pmatrix} p \\ q \end{pmatrix}
\mapsto \begin{pmatrix} p+\alpha \\ q \end{pmatrix} ,
\qquad
S_2^\alpha : \begin{pmatrix} p \\ q \end{pmatrix}
\mapsto \begin{pmatrix} p \\ q+\alpha \end{pmatrix} ,
$$
is explained in the following lemma.

\begin{lem} \label{t2}
For every $f\in\C^\infty(\TT^2)$, $a\in\ZZ$, we have (i) 
$$
t_1^{a} \Op_N(f) t_1^{-a} = \Op_N(f\circ S_2^{a/N}) ,
$$ 
$$
t_2^{a} \Op_N(f) t_2^{-a} = \Op_N(f\circ S_1^{-a/N}) ,
$$ 
and (ii) for all $\alpha\in\RR$,
$$
\|t_1^{a} \Op_N(f) t_1^{-a} - \Op_N(f\circ S_2^\alpha) \| 
\ll_f \big|\alpha-\frac{a}{N}\big|,
$$ 
$$
\| t_2^{a} \Op_N(f) t_2^{-a} - \Op_N(f\circ S_1^{-\alpha}) \|
\ll_f \big|\alpha-\frac{a}{N}\big|.
$$ 
\end{lem}

\begin{proof}
With the commutation relations (\ref{commrel}) we find
$$
t_1^{a} \Op_N(f) t_1^{-a} = \sum_{n} \widehat f(n) e_N(a n_2) T_N(n).
$$
On the other hand,
$$
\Op_N(f\circ S_2^\alpha) = \sum_{n} \widehat f(n) e(\alpha n_2) T_N(n),
$$
and the bound 
$$
\big| e(\alpha n_2) - e(\frac{a}{N} n_2) | \leq |2\pi n_2| 
\big|\alpha-\frac{a}{N}\big|
$$
concludes the proof of the statements concerning $t_1$.
The results for $t_2$ follow accordingly.
\end{proof}

\subsection{Friedrichs symmetrization}\label{fried}

Let $h\in{\mathcal S}(\RR^2)$ be an even, real-valued Schwartz function
normalized such that
$$
\int_{\RR^2} h(x)^2 dx=1 .
$$ 
The kernel
$$
K_N(x,x')= N^\frac12 \sum_{m\in\ZZ^2} h(N^\frac12 (x-x'+m))
$$
is now used to define an alternative quantization  (a variant of the
``anti-Wick quantization'') 
$$
\sOp_N(f) : \L^2(\ZZ_N) \rightarrow \L^2(\ZZ_N)
$$
of the observable $f\in\C^\infty(\TT^2)$, by setting
$$
\sOp_N(f) = \frac{1}{C_N}\int_{\TT^2} \big[ \Op_N(K_N(\,\cdot\, ,x)) \big]^2 
\; f(x)\,dx  .
$$
The normalization constant
$$
C_N= \sum_{n\in\ZZ^2} \int_{\RR^2} h(x)\, h(x+N^\frac12 n)\, dx 
$$
is chosen such that 
\begin{equation}
\sOp_N( 1 ) = \id_N .
\end{equation}
Asymptotically,
$$
C_N
= 1 + O_R(N^{-R}) ,  \quad \text{any $R$.}
$$

The main feature of this quantization is {\em positivity}: If $f\geq 0$ then 
\begin{equation}
\langle \sOp_N(f) \psi, \psi \rangle \geq 0 ,  
\end{equation}
since
$$
\langle \sOp_N(f) \psi, \psi \rangle  \\
=
\frac 1{C_N}\int_{\TT^2} 
\| \Op_N(K_N(\,\cdot\, ,x)) \psi \|_2^2 \;f(x) \,dx, 
$$
which is clearly non-negative. Hence 
$$
\mu_{N,\psi} : f \mapsto \langle \sOp_N(f) \psi, \psi \rangle
$$
defines a measure on $\C^\infty(\TT^2)$, with total mass
$\|\psi\|_2^2$.

This ``positive'' quantization
differs from the Weyl quantization at most by terms of order $1/N$, 
as stated in the following proposition.

\begin{prop}
For every $f\in\C^\infty(\TT^2)$ we have
$$
\| \Op_N(f) - \sOp_N(f) \| \ll_{f} \frac1N.
$$
\end{prop}

\begin{proof}
By the Poisson summation formula, our kernel $K_N$ can be re-expressed
in the form
$$
K_N(x,x')= \frac 1{N^{\frac 12}} \sum_{m\in\ZZ^2} 
\widehat h(\frac m{N^{\frac 12}})\, e(m(x-x')) ,
$$
where $\widehat h$ is the Fourier transform of $h$. Then, by definition,
$$
\Op_N(K_N(\,\cdot\, ,x)) = 
\frac 1{N^{\frac 12}} \sum_{m\in\ZZ^2} \widehat h(\frac m{N^{\frac 12}})\, 
e(-m x) \; T_N(m) 
$$
and
\begin{multline*}
\big[\Op_N(K_N(\,\cdot\, ,x)) \big]^2  
= 
\frac 1N \sum_{m,n\in\ZZ^2} \widehat h(\frac m{N^{\frac 12}}) \, 
\widehat h(\frac n{N^{\frac 12}})
\times \\ \times 
e(-(m+n) x) \,
e_N(\frac{\omega(m,n)}{2}) \; T_N(m+n) .
\end{multline*}
With this, we find
\begin{equation*}
\begin{split}
\sOp_N(f) &=
\frac{1}{N C_N} \sum_{k,m\in\ZZ^2} \widehat f(k)\, 
\widehat h(\frac m{N^{\frac 12}}) \, \widehat h(\frac {k-m}{N^{\frac 12}}) \,
e_N(\frac{\omega(m,k)}{2}) \; T_N(k) \\ 
& = \sum_{k\in\ZZ^2} \widehat f(k)  G_N(\frac k{N^{\frac 12}}) \TN(k)  
\end{split}
\end{equation*}
with 
$$
G_N(t) = \frac 1{NC_N} \sum_{m\in\ZZ^2}
\widehat h(\frac m{N^{\frac 12}}) \widehat h(t-\frac {m}{N^{\frac 12}}) \,
e(\frac12 \omega(\frac m{N^{\frac 12}},t)) .
$$
Therefore 
$$
\sOp_N(f) - \OPN(f) = \sum_{k\in\ZZ^2} \widehat f(k) 
\left( G_N(\frac k{N^{\frac 12}})-1\right )\TN(k)  .
$$

We have $G_N(0) = 1$ by Poisson summation and the definition of
$C_N$. 
It is easy to see that $G_N$ and its derivatives are bounded uniformly
in $N$ by rapidly decreasing functions of $t$. 
Moreover, $G_N(-t)=G_N(t)$ is {\em even} as is easy to see using $h$
is even and the bilinearity of $\omega$.  
Thus expanding $G_N(t)$ in a Taylor series at $t=0$ and noting that
since $G_N$ is even, the first order terms are  missing, we find that for 
$|t|\ll 1$,  
$$
G_N(t) = G_N(0) + O(|t|^2) = 1+O(|t|^2) .
$$
Therefore since the Fourier coefficients
$\widehat f(k)$ are rapidly decreasing,
\begin{equation*}
\begin{split}
\norm{\sOp_N(f) - \OPN(f)} &\ll \sum_{k\in\ZZ^2} |\widehat f(k) | 
| G_N(\frac k{N^{\frac 12}}) -1 |  \\
&\ll \sum_{|k| \leq N^{1/10}} |\widehat f(k) | \frac{|k|^2}N + 
 \sum_{|k| > N^{1/10}}  |\widehat f(k) | \\
&\ll_f \frac 1N + \frac 1{N^R} \ll \frac 1N ,
\end{split}
\end{equation*}
and the Proposition follows.  
\end{proof}

\newpage
\section{Quantizing skew translations}

In this section we define a quantization 
$\UaN: \L^2(\ZZ_N) \rightarrow \L^2(\ZZ_N)$ 
for the skew translation of the torus
$$
A_\alpha : \begin{pmatrix} p \\ q \end{pmatrix}
\mapsto \begin{pmatrix} p +\alpha \\ q + 2 p \end{pmatrix} .
$$
We define the quantization 
in  the momentum representation, that is $\UaN = \FN^{-1} \VaN \FN$, by 
choosing an approximation $a/N$ to $\alpha$, with 
\begin{equation}\label{alpha -a/N}
|\alpha -\frac aN|<\frac 1N
\end{equation}
and then setting
$$
[\VaN \psi](P) := e_N(-(P-a)^2)\, \psi(P-a) .
$$

The relation between the quantized map and the classical map
$A_\alpha$ is given by  
\begin{thm}[Egorov's Theorem for $A_\alpha$]\label{egorov thm}
If $|\alpha-a/N|<1/N$ then for every $f\in\C^\infty(\TT^2)$ we have 
$$
\big\| \UaN^{-1}\, \Op_N(f) \,\UaN - \Op_N(f\circ A_\alpha) \big\|
\ll_f \frac 1N .
$$
\end{thm}
This is an immediate conclusion of the following Proposition 
together with the choice \eqref{alpha -a/N}:
\begin{prop}
For every $f\in\C^\infty(\TT^2)$ we have (i) 
\begin{equation}\label{exact Egorov}
\UaN^{-1}\, \Op_N(f) \,\UaN = \Op_N(f\circ A_{a/N})  ,
\end{equation}
and (ii) for all $\alpha\in \RR$,
$$
\big\| \UaN^{-1}\, \Op_N(f) \,\UaN - \Op_N(f\circ A_\alpha) \big\|
\ll_f \big| \alpha -\frac{a}{N} \big| .
$$
\end{prop}
\begin{proof}
We have to show that 
$$
\VaN^{-1}\, \widehat\Op_N(f) \,\VaN = \widehat\Op_N(f\circ A_{a/N})  ,
$$
and that for real $\alpha$, 
$$
\big\| \VaN^{-1}\,\widehat\Op_N(f) \,\VaN - 
\widehat\Op_N(f\circ A_\alpha) \big\|
\ll_f \big| \alpha -\frac{a}{N} \big| 
$$
where
$$
\widehat\Op_N(f) = \FN\Op_N(f)\FN^{-1} .
$$
Note that we can write 
$$
\VaN = t_1^{-a} \, \VoN , \qquad A_\alpha = S_1^\alpha \circ A_0 .
$$
Since 
\begin{equation*}
\begin{split}
\VaN^{-1}\,\widehat\Op_N(f) \,\VaN 
&=\VoN^{-1} t_1^a \FN\Op_N(f)\FN^{-1} t_1^{-a} \VoN \\
&=\VoN^{-1} \FN t_2^{-a} \Op_N(f) t_2^{a} \FN^{-1} \VoN ,
\end{split}
\end{equation*}
we find, by virtue of Lemma \ref{t2},
$$
\VaN^{-1}\,\widehat\Op_N(f) \,\VaN = 
\VoN^{-1}\, \widehat\Op_N(f\circ S_1^{a/N})\,\VoN ,
$$
and
$$
\big\| \VaN^{-1}\,\widehat\Op_N(f) \,\VaN - 
\VoN^{-1}\, \widehat\Op_N(f\circ S_1^\alpha)\,\VoN \big\|  
\ll_f \big| \alpha -\frac{a}{N} \big| ,
$$
respectively. It thus remains to be checked that
$$
\VoN^{-1}\, \widehat\Op_N(f)\,\VoN 
= \widehat\Op_N(f\circ A_0) .
$$
To this end, note first that
$$
\widehat\Op_N(f) 
= \sum_n \widehat f(n) e_N(\frac{n_1 n_2}{2}) t_1^{-n_2} t_2^{n_1} .
$$
Second, let us show that 
\begin{equation}\label{cmrel}
\VoN^{-1}\,t_1^{-n_2} t_2^{n_1} \,\VoN
=e_N( n_2^2)\, t_1^{-n_2} t_2^{n_1+2n_2} 
\end{equation}
holds:
\begin{equation*}
\begin{split}
[\VoN^{-1}\,t_1^{-n_2} & t_2^{n_1} \,\VoN \psi](P) \\
& = e_N(P^2) [t_2^{n_1} \,\VoN ] \psi(P-n_2) \\
& = e_N(P^2+n_1(P-n_2)-(P-n_2)^2) \psi(P-n_2) \\
& = e_N(n_2^2) e_N\big((n_1+2n_2)(P-n_2)\big) \psi(P-n_2) \\
& = e_N(n_2^2) [t_2^{n_1+2n_2} \psi] (P-n_2) \\
& = e_N(n_2^2) [t_1^{-n_2} t_2^{n_1+2n_2} \psi] (P) .
\end{split}
\end{equation*}
The commutation relations (\ref{cmrel}) now lead to
$$
\VoN^{-1}\, \widehat\Op_N(f)\,\VoN 
=\sum_n \widehat f(n_1-2 n_2, n_2) e_N(\frac{n_1 n_2}{2}) t_1^{-n_2} t_2^{n_1} .
$$
The Fourier coefficients
of $f \circ A_0$ are, however, exactly $\widehat f(n_1-2 n_2, n_2)$,
and our proof is complete.
\end{proof}

\begin{remark} There are other quantization schemes of skew
translations in the literature \cite{DeBievre96,Bouzouina96}. 
However, they do not satisfy 
Theorem~\ref{egorov thm} and so their relevance to the classical dynamics is
unclear. 
\end{remark}
\subsection{Proof of Theorem~\ref{abs thm}} 
For each $N$, choose an approximant $a/N$ with $|\alpha -a/N|\to 0$,
and a normalized eigenfunction  $\psi\in \HN$ of $\UaN$. 
Using the results of Section \ref{fried}, we get a sequence of {\em probability
measures} $\mu_{N,\psi}(f) = \langle \sOp_N(f)\psi,\psi \rangle$.  
Since they differ from the distributions 
$f\mapsto \langle \OPN(f)\psi,\psi \rangle$ by terms 
which vanish as $N\to \infty$, it suffices to show that
$\mu_{N,\psi}$ converge to Lebesgue measure 
$\lambda :f\mapsto \int_{\TT^2}f$.  

To see this, note that the space of probability measures on $\TT^2$ is
compact, and hence any sequence of probability measures has a
convergent subsequence. Thus the sequence $\mu_{N,\psi}$ has a limit
point, which is  a probability measure.  
Any such limit point $\nu$ is then invariant under the map $A_\alpha$
by Egorov's theorem (Theorem~\ref{egorov thm}). 
For irrational $\alpha$, the map $A_\alpha$ is
uniquely ergodic which forces $\nu=\lambda$. Thus Lebesgue measure 
$\lambda$ is the unique accumulation point of our sequence. 
This forces $\mu_{N,\psi}\to \lambda$, otherwise there would be a
neighborhood of $\lambda$ which excludes infinitely many
$\mu_{N,\psi}$. But then these latter would have to contain a
convergent subsequence whose limit would not be $\lambda$ --- 
a contradiction.

\newpage 
\section{Upper bounds for the rate of quantum unique ergodicity}

Besides the convergence result of Theorem \ref{abs thm}, 
we can also give a bound for the rate of convergence. 
To do this, we will always assume that we pick approximants such that 
$|\alpha-a/N|<1/N$. 
Our first result is 
\begin{thm}\label{thm upper} 
If $\alpha$ is irrational then for all $f\in \C^\infty(\TT^2)$, and any
normalized eigenfunction $\psi\in L^2(\ZZ_N)$ of the propagator $\UaN$,  
$$
\langle \OPN(f)\psi,\psi \rangle =\int_{\TT^2}f(p,q)\, dp\, dq 
+ O_f( \frac 1{M^{1/2}}),\quad N\to \infty ,
$$
where $M=N/\gcd(a,N)$. 
\end{thm}
To see that this has content, we note
\begin{lem}\label{M grows with N}
Suppose we take approximations $a/N$ to $\alpha$ with  
$\frac aN \to\alpha $ as $N\to\infty$. 
If $\alpha$ is irrational then
$N/\gcd(a,N)\to\infty$ as $N\to\infty$. 
\end{lem}
Indeed, write $D=\gcd(a,N)$, $M=N/D$, $b=a/D$. 
If $\alpha$ is irrational then 
$c_M:=\min\{|\alpha-\frac km| : m\leq M\}>0$,  and so 
$ |\alpha -\frac aN| = |\alpha -\frac bM| \geq c_M>0$.  
Thus if $M$ is bounded then  $c_M$ is bounded away from zero, 
contradicting $\frac aN \to\alpha $.

We say that an irrational $\alpha$ is {\em badly approximable} if 
$$
|\alpha- \frac an| \gg_\epsilon 
\frac 1{n^{2+\epsilon}},\quad \forall \epsilon >0 .
$$
In that case, we can say something stronger then just that $M\to
\infty$ as $N\to \infty$. In fact we have $M\gg_\epsilon N^{1/2-\epsilon}$
since 
$$
\frac 1N > |\alpha-\frac aN|=|\alpha-\frac bM| 
\gg_\epsilon \frac 1{M^{2+\epsilon}} .
$$
We thus find: 
\begin{cor}\label{cor badly app}
If $\alpha$ is badly approximable then for any normalized
eigenfunction $\psi$ of the propagator $\UaN$ we have
$$ 
\langle \OPN(f)\psi,\psi \rangle -\int_{\TT^2}f(p,q)\, dp\, dq 
\ll_{f,\epsilon} N^{-1/4+\epsilon} \quad\text{for all $\epsilon>0$.}
$$
\end{cor}

\subsection{Proof of Theorem \ref{thm upper}} 
The idea of the proof of Theorem~\ref{thm upper} is to use the fact 
that for an eigenfunction $\psi$ of $\UaN$, we have 
$\langle \OPN(f) \psi,\psi\rangle = 
\langle \OPN(f^T)\psi,\psi\rangle$, 
where $f^T:=\frac 1T\sum_{t=1}^T f\circ A_{a/N}^t$ is the ergodic
average of $f$. Taking $T=N$ we show directly that for {\em any}
$\psi\in \HN$, we have 
$|\langle \OPN(f^N)\psi,\psi\rangle-\int_{\TT^2} f|\ll M^{-1/2}$. 

We start the argument by taking for $f$ the basic exponential
$e_{m,n}(p,q):=e(mp+nq)$.  
\begin{lem}\label{lemT(n)}
For any normalized $\psi\in L^2(\ZZ_N)$, 
\begin{equation}
|\langle \OPN(e_{m,n}^T) \psi,\psi\rangle |^2 \leq 
\frac 1{T^2 N} \sum_{y\bmod N} |\widehat\psi(y)|^2
|S(\frac {an}N, \frac{a(m-n)+n^2+2ny}N;T)|^2 ,
\end{equation}
where 
\begin{equation}
S(a,b;T) := \sum_{t=1}^T e(at^2+bt) .
\end{equation}
\end{lem}
\begin{proof}
Iterating $A_\alpha$   gives 
\begin{equation}\label{iterate}
A_\alpha^t \begin{pmatrix}p\\ q\end{pmatrix} = 
\begin{pmatrix} p+t\alpha\\ q+2tp+t(t-1)\alpha \end{pmatrix}\bmod 1 .
\end{equation}
 From \eqref{iterate} we find that 
\begin{equation}\label{basic exponential}
e_{m,n}\circ A_{a/N}^t = e_N(ant^2+a(m-n)t)) e_{m+2tn,n} .
\end{equation}
 Therefore
$$
e_{m,n}^T:=\frac 1T\sum_{t=1}^T e_{m,n}\circ A_{a/N}^t = 
\frac 1T\sum_{t=1}^T e_N(ant^2+a(m-n)t)) e_{m+2tn,n} ,
$$
and quantizing we get 
\begin{equation}\label{OPN(e_{m,n}^T)}
\OPN(e_{m,n}^T)  = \frac 1T\sum_{t=1}^T e_N(ant^2+(m-n)t)) \TN(m+2tn,n) .
\end{equation}
In particular, $\norm{\OPN(e_{m,n}^T)} \leq 1$.

 From \eqref{OPN(e_{m,n}^T)} we get 
\begin{multline*}
\OPN(e_{m,n}^T) \psi(x) =\\ 
\frac 1T \sum_{t=1}^T e_N\left(ant^2 +a(m-n)t + \frac 12 (m+2tn)n +nx\right) 
\psi(x+m+2tn) ,
\end{multline*}
and so 
\begin{multline*}
\langle \OPN(e_{m,n}^T) \psi,\psi\rangle  = 
\frac{e_N(\frac{mn}2)}{TN} \sum_{x\bmod N} \overline{\psi(x)}e_N(nx) \\
\cdot \sum_{t=1}^T e_N \left( ant^2+(a(m-n)+n^2)t \right)\psi(x+m+2tn) .
\end{multline*}
On applying Cauchy-Schwarz we find
\begin{multline*}
|\langle \OPN(e_{m,n}^T) \psi,\psi\rangle |^2 \leq 
\frac 1{T^2} \frac 1N\sum_{x\bmod N} |\psi(x)|^2 \\ 
\cdot \frac 1N\sum_{x\bmod N}|
\sum_{t=1}^T e_N \left( ant^2+(a(m-n)+n^2)t \right) \psi(x+m+2tn)|^2 .
\end{multline*}
Now $\frac 1N\sum_{x\bmod N}|\psi(x)|^2 =\norm{\psi}^2=1$ 
and using the Fourier expansion
$\psi(x)=N^{-1/2} \sum_y \widehat\psi(y) e(yx)$ in the second $x$-sum
gives 
\begin{equation*}
\begin{split}
|\langle & \OPN(e_{m,n}^T) \psi,\psi\rangle |^2  \\
&\leq 
\frac 1{T^2} 
\frac 1{N^2}\sum_{x,y,y'\bmod N} \widehat\psi(y)\overline{\widehat\psi(y')}
\;e_N(x(y-y')) \\
& \cdot
\sum_{t,t'=1}^T e_N(an(t^2-{t'}^2)+(a(m-n)+n^2)(t-t')+2nyt-2ny't') \\
&= \frac 1{T^2N}\sum_{y\bmod N}|\widehat\psi(y)|^2 \cdot
 |\sum_{t=1}^T e_N(ant^2+(a(m-n)+n^2+2ny)t) |^2 ,
\end{split}
\end{equation*}
by Parseval's identity.
\end{proof}
We now take $T=N$ and then get a  Gauss sum for $S(a,b;T)$ in
Lemma~\ref{lemT(n)}: 
Define the complete Gauss sum 
$$
G(c,d;N):=\sum_{t\bmod N}e_N(ct^2+dt) .
$$
We will need a very classical estimate of its absolute value, which we
recall:  
\begin{lem}\label{gauss sum}
If $\gcd(2c,N)=1$ then 
$$
|G(c,d;N)| =N^{1/2} .
$$
If $c\neq 0\bmod N$ then 
$$
|G(c,d;N)|\leq N^{1/2}\gcd(2c,N)^{1/2},\qquad c\neq 0 \bmod N .
$$
If $c=0\bmod N$ then 
$$
G(0,d;N) = \begin{cases} N,&d\equiv 0 \bmod N\\0,& d\neq 0\bmod N .
\end{cases} 
$$
\end{lem} 
\begin{proof}
Since the case $c=0\bmod N$ is obvious, we assume $c\neq 0\bmod N$. 
By multiplying out $|G(c,d;N)|^2$, changing variables and switching 
the order of summation we find 
\begin{equation*}
\begin{split}
|G(c,d;N)|^2&= \sum_{t_1,t_2\bmod N}e_N(c(t_1^2-t_2^2)+d(t_1-t_2))\\
& = \sum_{y\bmod N}e_N(cy^2+dy)\sum_{t\bmod N}e_N(2cyt) .
\end{split}
\end{equation*}
The inner sum is either $N$ or $0$, depending if $2cy=0\bmod N$ or
not. This gives 
$$
|G(c,d;N)|^2= N\sum_{y: 2cy=0\bmod N} e_N(cy^2+dy) .
$$
If $\gcd(2c,N)=1$ then the only solution of $2cy=0\bmod N$ is 
$y=0\bmod N$ so we get equality $|G(c,d;N)|^2= N$, while in general the
number of solutions is $\gcd(2c,N)$ which gives the bound $|G(c,d;N)|^2
\leq N\gcd(2c,N)$. 
\end{proof}

\begin{lem}\label{lemT(n)2}
For any normalized $\psi\in L^2(\ZZ_N)$, and $|m|,|n|<M$, $(m,n)\neq (0,0)$,
we have, if $n\neq 0$,
\begin{equation}
|\langle \OPN(e_{m,n}^N) \psi,\psi\rangle | 
\leq |2n|^{1/2} M^{-1/2} ,
\end{equation}
while if $n=0$ but $m\neq 0\mod M$ then 
\begin{equation}
\langle \OPN(e_{m,0}^N) \psi,\psi\rangle  =0  .
\end{equation}
\end{lem}
\begin{proof}
 From Lemma~\ref{lemT(n)} we have 
$$
|\langle \OPN(e_{m,n}^N) \psi,\psi\rangle |^2 \leq 
\frac 1{N^3} \sum_{y\bmod N} |\widehat\psi(y)|^2 |G(an, a(m-n)+n^2+2ny;N)|^2 .
$$
Recall that $a/N=b/M$ with $b,M$ co-prime, $D=\gcd(a,N)$. 
We have $an\neq 0 \bmod N$ if and only if $n\neq 0\bmod M$. 
Thus if $n\neq 0\bmod M$
then by Lemma~\ref{gauss sum} 
\begin{equation*}
\begin{split}
|\langle \OPN(e_{m,n}^N) \psi,\psi\rangle |^2 &\leq
N^{-2} \sum_{y\bmod N} |\widehat\psi(y)|^2\, \gcd(2an,N)\\ 
&= M^{-1} \gcd(2n,M) \|\widehat\psi\|^2 \\
&\leq |2n| M^{-1} ,
\end{split}
\end{equation*}
since $\|\widehat\psi\|=\|\psi\|=1$. 

If $n=0$ then by \eqref{basic exponential}, 
$$
e^N_{m,0} = e_{m,0}\cdot \frac 1N \sum_{t\bmod N} e_N(amt)
$$
which vanishes if $am\neq 0\mod N$, equivalently if $m\neq 0\mod
M$. Thus $\OPN(e^N_{m,0})=0$ if $m\neq 0\mod M$. 
%\begin{multline*}
%|\langle \OPN(e_{m,n}^N) \psi,\psi\rangle |^2 \leq \frac 1{N^3}
%\sum_{y\bmod N} |\widehat\psi(y)|^2\, |G(0,am;N)|^2 \\ 
%=  
%\begin{cases} 1,& am=0 \bmod N\\ 0& \text{ otherwise,}
%\end{cases}
%\end{multline*}
%the above condition being equivalent to $m=0 \bmod M$.
\end{proof}

\subsection{Conclusion of the proof} 
If $\psi$ is an eigenfunction of $\UaN$, then 
\begin{equation*}
\begin{split}
\langle \OPN(f)\psi,\psi \rangle &= 
\frac 1T \sum_{t=1}^T 
\langle \OPN(f)\UaN^t\psi,\UaN^t\psi \rangle
\\ 
&= 
\frac 1T \sum_{t=1}^T 
\langle \UaN^{-t}\OPN(f)\UaN^t\psi,
\psi\rangle .
\end{split}
\end{equation*}
By Egorov \eqref{exact Egorov},
$$
\UaN^{-t}\OPN(f)\UaN^t=
\OPN(f\circ A_{a/N}^t)
$$ 
and so 
$$
\langle \OPN(f)\psi,\psi \rangle = 
\langle \OPN(f^T)\psi,\psi \rangle .
$$

Then expanding $f$ in a Fourier series 
$f=\sum_{(m,n)}\widehat f(m,n) e_{m,n}$ and applying the
ergodic average operator with $T=N$ we get 
$$
f^N=\sum_{(m,n)\neq (0,0)}\widehat f(m,n) e_{m,n}^N .
$$
Therefore 
$$
\langle \OPN(f^N)\psi,\psi \rangle  -\int_{\TT^2}f(p,q)\, dp\, dq  
= \sum_{(m,n)\neq (0,0)}
\widehat f(m,n) \langle \OPN(e_{m,n}^N) \psi,\psi\rangle .
$$ 
Now we have $\norm{\OPN(e_{m,n}^N)}\leq 1$ and so we truncate the sum
above to frequencies $|m|,|n|<M$ with error at most 
$$
\sum_{\substack{|m|\geq M\text{ or }|n|\geq M\\(m,n)\neq(0,0)}} 
|\widehat f(m,n)| \ll M^{-K} 
$$
since $\widehat f(m,n)$ is rapidly decreasing. 
It is important to note that since $\alpha$ is irrational, we have
$M\to \infty$ as $N\to\infty$ which we assume.  

For the small
frequencies we use Lemma~\ref{lemT(n)2} to find 
\begin{equation*}\begin{split}
\sum_{\substack{|m|,|n|<M\\(m,n)\neq (0,0)}}
\widehat f(m,n) \langle \OPN(e_{m,n}^N) \psi,\psi\rangle
&\ll \sum_{\substack{|m|,|n|<M\\ n\neq 0}} 
|\widehat f(m,n)| |n|^{1/2} M^{-1/2}\\ 
&\ll_f M^{-1/2} .
\end{split}\end{equation*}
Thus we find that for normalized eigenfunctions $\psi$ we have 
$$
\langle \OPN(f)\psi,\psi \rangle -\int_{\TT^2}f(p,q)\, dp\, dq  \ll_f
M^{-1/2} +M^{-K}\ll M^{-1/2} .
$$
This concludes the proof of Theorem~\ref{thm upper}.  
\newpage
\section{Explicit eigenfunctions}\label{explicit sec}

We begin by calculating the eigenvalues and a basis of eigenfunctions 
of the quantum map $\VaN$, defined by
$$
[\VaN \psi](P) =  e_N(-(P-a)^2)\,\psi(P-a) .
$$
The eigenvalue equation 
\begin{equation}\label{ev}
\VaN\; \psi = e_N(\phi)\; \psi
\end{equation}
yields the following simple recursion relation
for the eigenfunction $\psi$,
\begin{equation}\label{rec}
\psi(P+a)= e_N(-\phi-P^2)\, \psi(P).
\end{equation}
We can now construct 
$N$ linearly independent solutions $\psi_j$ of
(\ref{ev}), $j=1,\ldots,N$, as follows.
Let $D=\gcd(a,N)$ be the greatest common divisor of
$a$ and $N$. Put 
$$
b=\frac{a}{D}, \quad M=\frac{N}{D} ,
$$
and write furthermore
$$
j=\eta + D l, \quad \eta\in[1, D], \quad l\in[0,M-1].
$$
For a given $j\in[1,N]$, the pair $(\eta,l)$ is uniquely determined.

\begin{prop}
The functions
$$
\psi_{\eta,l}(P)=
\begin{cases}
\sqrt{D}\, e_N\big( -\eta a \nu^2 - \nu l D 
+ a^2 \nu \frac{(M-1)(2M-1)-(\nu-1)(2\nu-1)}{6} \big) & \\
\hskip170pt \text{if $P\equiv \eta + \nu a \bmod N$} & \\ 
0 & \\
\hskip170pt \text{if $P\not\equiv \eta \bmod D$}\end{cases}
$$
solve Equation (\ref{ev}) with eigenphases
\begin{equation}\label{eigenphases}
\phi_{\eta,l} = l D - \eta^2 + \eta a -  a^2 \frac{(M-1)(2M-1)}{6},
\end{equation}
and form an orthonormal basis of $\L^2(\ZZ_N)$.
\end{prop}

\begin{proof}
Put $\psi_{\eta,l}(\eta)=\sqrt{D}$. The recursion relation (\ref{rec}) then
implies that $\psi_{\eta,l}$ at points of the form $P=\eta+\nu a$ 
($\nu=1,2,\ldots$) reads
$$
\psi_{\eta,l}(P)=
\sqrt{D} \, e_N\big(-\nu \phi_{\eta,l} -\sum_{m=0}^{\nu-1} (\eta+m a)^2 \big).
$$
Since $N$ divides $Ma$, and thus
$\psi_{\eta,l}(\eta+Ma)=\psi_{\eta,l}(\eta)$, the eigenphases are determined
by
$$
e_N\big( M \phi_{\eta,l} + \sum_{m=0}^{M-1} (\eta+m a)^2 \big) = 1,
$$
leaving an ambiguity mod $\frac{N}{M}$, which permits to put
\begin{multline*}
\phi_{\eta,l}= l D - \frac{1}{M} \sum_{m=0}^{M-1} (\eta+m a)^2  \\
= l D - \eta^2 - \eta a (M-1) -  a^2 \frac{(M-1)(2M-1)}{6} .
\end{multline*}
Since $\eta a M \equiv 0$ mod $N$, we drop this term.
A straightforward manipulation leads to the expression for $\psi_{\eta,l}$
as given in the proposition. Orthonormality follows from
\begin{equation*}
\begin{split}
\langle \psi_{\eta,l},\psi_{\eta',l'} \rangle
& = \frac{1}{N} \sum_{P\bmod N} \psi_{\eta,l}(P) \overline\psi_{\eta',l'}(P) \\
& = 
\begin{cases}
\displaystyle
\frac{D}{N} \sum_{\nu=0}^{M-1} e_N\big(-\nu (l-l') D \big) &
\text{if $\eta=\eta'$} \\
0 & \text{if $\eta \neq \eta'$} 
\end{cases}
\end{split}
\end{equation*}
and
$$
\frac{1}{M} \sum_{\nu=0}^{M-1} e_M\big(\nu (l-l') \big) 
=
\begin{cases}
1 & \text{if $l=l'$} \\
0 & \text{otherwise.} 
\end{cases}
$$
\end{proof}

\begin{cor}
The multiplicity $m(\phi)$ of an eigenphase $\phi$ is bounded by 
$m(\phi) \ll D^\frac12 \tau(D) \ll_\epsilon D^{\frac12+\epsilon}$,
any $\epsilon>0$, where $\tau(D)$ is the number of divisors of $D$.
\end{cor}

\begin{proof}
For a given $\phi$, we would like to count the number of solutions
of
\begin{equation}\label{count}
l D - \eta^2 + \eta a -  a^2 \frac{(M-1)(2M-1)}{6} = \phi \bmod N.
\end{equation}
This implies that $-\eta^2 \equiv \phi \bmod D$. In order to count
the number $\#_{D,\phi}$ of solutions of the latter equation, define
$$
\delta_D(x)=
\begin{cases}
1 & \text{if $x\equiv 0 \bmod D$} \\
0 & \text{if $x\not\equiv 0 \bmod D$} .
\end{cases}
$$
Then
$$
\#_{D,\phi} = \sum_{P=1}^D \delta_D(P^2+\phi) .
$$
Since 
$$
\delta_D(x) = \frac1D \sum_{\rho\bmod D} e_D(\rho x)
$$
we find that
$$
\#_{D,\phi} = \frac1D \sum_{\rho\bmod D} e_D(\rho \phi)
\bigg(\sum_{P=1}^D e_D(\rho P^2) \bigg) .
$$
The sum in brackets is a classical Gauss sum, whose absolute value is bounded
by $\sqrt{D \gcd(D,2\rho)}$ (Lemma~\ref{gauss sum}), and thus
$$
\#_{D,\phi} \ll D^\frac12 \tau(D).
$$
For fixed $\eta$, Equation (\ref{count}) determines $l$ uniquely mod $M$. 
\end{proof}

Let us put
$$
m_2= -n_1 b , \qquad m_1= (n_2 -2\mu \eta - \mu(\mu-1)a) b - (l-l') .
$$

\begin{lem} \label{lemT}
We have 
$$
|\langle T_N(n) \psi_{\eta,l}, \psi_{\eta',l'} \rangle| 
= \begin{cases} 
\frac {| \sum_{\nu=0}^{M-1} e_M(m_2 \nu^2 + m_1 \nu) |}M ,
&\eta-\eta'=n_1\bmod D\\0,& \text{ otherwise} 
\end{cases}
$$
and in particular for $n_1 \equiv 0 \bmod M$
$$
|\langle T_N(n) \psi_{\eta,l}, \psi_{\eta',l'} \rangle |
= \begin{cases}
0 & \text{if $m_1 \not\equiv 0 \bmod M$} \\
1 & \text{if $m_1 \equiv 0 \bmod M$.}
\end{cases}
$$
\end{lem}

\begin{proof}
We have
\begin{multline*}
\langle T_N(n) \psi_{\eta,l}, \psi_{\eta',l'} \rangle
= \frac{1}{N} \sum_{P\bmod N} e_N(\frac{n_1 n_2}{2} + n_2 P)\,
\psi_{\eta,l}(P+n_1)\, \overline\psi_{\eta',l'}(P) \\
=\frac{1}{N} \sum_{\nu=0}^{M-1} e_N(\frac{n_1 n_2}{2} + n_2 (\eta'+\nu a))\,
\psi_{\eta,l}(\eta'+\nu a +n_1)\, \overline\psi_{\eta',l'}(\eta'+\nu a) ,
\end{multline*}
which is non-zero only if there is a $\mu$ such that
\begin{equation}\label{n1}
n_1\equiv \mu a +\eta - \eta' \bmod N, 
\end{equation}
hence in particular
$n_1+\eta'-\eta \equiv 0$ mod $D$.
Taking absolute values and using the explicit expressions for the
eigenfunctions we obtain
\begin{multline} \label{dfg}
|\langle T_N(n) \psi_{\eta,l}, \psi_{\eta',l'} \rangle| \\
= \frac{1}{M} \bigg| \sum_{\nu=0}^{M-1} 
e_N\bigg[ n_2\nu a - (\phi_{\eta,l}-\phi_{\eta',l'})\nu 
- \sum_{m=0}^{\nu+\mu-1} (\eta+m a)^2 
+ \sum_{m=0}^{\nu-1} (\eta'+m a)^2 \bigg]  \bigg| .
\end{multline}
By virtue of Relation (\ref{n1}) we have 
$$
\sum_{m=0}^{\nu-1} (\eta'+m a)^2
\equiv \sum_{m=\mu}^{\nu+\mu-1} (\eta+m a-n_1)^2 \bmod N .
$$
This formula allows us to simplify (\ref{dfg}) to
\begin{multline} \label{dfgg}
|\langle T_N(n) \psi_{\eta,l}, \psi_{\eta',l'} \rangle| \\
= \frac{1}{M} \bigg| \sum_{\nu=0}^{M-1} 
e_N\bigg[ n_2\nu a - (\phi_{\eta,l}-\phi_{\eta',l'})\nu - n_1 a \nu (\nu-1)
- n_1(2\mu a + 2 \eta - n_1) \nu \bigg]  \bigg| \\
= \frac{1}{M} \bigg| \sum_{\nu=0}^{M-1} 
e_M( m_2 \nu^2 + m_1 \nu) \bigg| ,
\end{multline}
with
$$
m_2= -n_1 b , \qquad m_1= (n_2 -2\mu \eta - \mu(\mu-1)a) b - (l-l') .
$$
In the case $n_1 \equiv 0 \bmod M$ we have $m_2\equiv 0 \bmod M$
and thus
$$
|\langle T_N(n) \psi_{\eta,l}, \psi_{\eta',l'} \rangle |
=\frac{1}{M} \bigg| \sum_{\nu=0}^{M-1} e_M(m_1\nu) \bigg|
= \begin{cases}
0 & \text{if $m_1 \not\equiv 0 \bmod M$} \\
1 & \text{if $m_1 \equiv 0 \bmod M$.}
\end{cases}
$$
\end{proof}

In the sequel $a\in\ZZ$ will be chosen such that
$$
|\alpha-\frac{a}{N}|<\frac1N
$$
holds.

\begin{prop} \label{diag}
Let $f\in\C^\infty(\TT^2)$, and assume $\alpha$ is diophantine. Then
$$
\langle \Op_N(f) \psi_{\eta,l}, \psi_{\eta,l} \rangle
= \int_{\TT^2} f\, dx 
+O_{f,\alpha} (N^{-\frac12}) .
$$
If $f$ is a polynomial, then the above relation holds for all irrational
$\alpha$.
\end{prop}

\begin{proof}
Without loss of generality we assume $\int_{\TT^2} f\, dx=0$.
We have
$$
\langle \Op_N(f) \psi_{\eta,l}, \psi_{\eta,l} \rangle
= \sum_{n\neq 0} \widehat f(n) \langle T_N(n) \psi_{\eta,l}, \psi_{\eta,l} \rangle,
$$
where $\widehat f(n)$ are the (rapidly decreasing) Fourier coefficients of $f$.

Following Lemma \ref{lemT}, we distinguish two cases.

{\em Case A:} $n_1\not\equiv 0\bmod M$. Then 
$m_2=-ab\mu \equiv -b n_1\not\equiv 0\bmod M$ and by Lemma~\ref{gauss sum}
$$
\bigg| \sum_{\nu=0}^{M-1} e_M(m_2 \nu^2 + m_1 \nu) 
\bigg| \leq \sqrt{M \gcd(M,2 |n_1|)} \leq  \sqrt{2 |n_1| M},
$$
from which we obtain
$$
|\langle T_N(n) \psi_{\eta,l}, \psi_{\eta,l} \rangle | \leq \sqrt{2|n_1|} M^{-\frac12} .
$$ 

{\em Case B.} $n_1 \equiv 0 \bmod M$.
Here
$$
|\langle T_N(n) \psi_{\eta,l}, \psi_{\eta,l} \rangle |
=\frac{1}{M} \bigg| \sum_{\nu=0}^{M-1} e_M(n_2 b \nu) \bigg|
= \begin{cases}
0 & \text{if $n_2 \not\equiv 0 \bmod M$} \\
1 & \text{if $n_2 \equiv 0 \bmod M$.}
\end{cases}
$$

In summary,
\begin{equation}\label{summ}
|\langle \Op_N(f) \psi_{\eta,l}, \psi_{\eta,l} \rangle |
\leq \sqrt2 M^{-\frac12} \sum_{\substack{ n_1 \neq 0 \\ D| n_1}} 
|n_1|^\frac12 | \widehat f(n) | + 
\sum_{\substack{n\neq 0 \\ D | n_1 \\ M|n_1,n_2}} | \widehat f(n) | ,
\end{equation}
the first sum corresponds to Case A, and the second sum to Case B.
Since $f$ is smooth we can bound the first sum by
$$
\sqrt2 M^{-\frac12} \sum_{\substack{ n_1 \neq 0 \\ D| n_1}} 
|n_1|^\frac12 |\widehat f(n) | 
\ll_{f,R} M^{-\frac12} D^{-R}  = N^{-\frac12} D^{-R+\frac12}
$$
for any $R$. Hence
$$
\sqrt2 M^{-\frac12} \sum_{\substack{ n_1\neq 0 \\ D| n_1}} 
|n_1|^\frac12 |\widehat f(n) | 
\ll_{f} N^{-\frac12}  .
$$
The second sum in (\ref{summ}) is bounded by 
$$
\sum_{\substack{n\neq 0 \\ M|n_1,n_2}} | \widehat f(n) |  
\ll_{f,R} M^{-R}  .
$$
In particular the sum is empty for $N$ large enough,
if $f$ is a polynomial, because for
$\alpha$ irrational, $M$ grows with $N$ (see Lemma~\ref{M grows with N}). 
Thus we get the second part of the proposition. 
\end{proof}

\newpage
\section{Lower bounds}
We begin with a result which implies that our bound on the rate of
convergence (Corollary~\ref{cor badly app}) for badly 
approximable $\alpha$ is the optimal one:  
\begin{thm}\label{thm lower all}
For any irrational $\alpha$, there are arbitrarily large $N$, 
approximants $|\alpha -a/N|<1/N$  and eigenfunctions $\psi$ of $\UaN$ 
so that 
$$
|\langle \TN(2,0)\psi,\psi \rangle|  = \frac 1{2N^{1/4}} .
$$
\end{thm}

Unlike badly approximable $\alpha$'s, where we have an upper bound
on the rate of convergence of $1/N^{1/4}$ (Corollary~\ref{cor badly app}), 
we can construct irrationals
for which the rate of convergence  is arbitrarily
slow, e.g. slower then $1/\log\log\log N$: 
\begin{thm}\label{thm lower} 
Let $g(x)$ be an increasing positive function.  
Then there is an irrational $\alpha$ such that  
there are arbitrarily large values of $N$ for which there are 
normalized eigenfunctions $\psi$ of $\UaN$ satisfying
$$
|\langle \TN(2,0)\psi,\psi \rangle| \gg \frac 1{g(N)} .
$$
\end{thm}

\subsection{Constructing special eigenfunctions}  
In order to prove Theorems \ref{thm lower all} and \ref{thm lower}, 
we first use the results of Section~\ref{explicit sec} to construct special eigenfunctions 
$\psi$ for which the upper bound of Theorem~\ref{thm upper} is
optimal:  
\begin{prop}\label{special eigen} 
If $D=\gcd(a,N)>2$ and $M=N/\gcd(a,N)$ is odd then there are normalized 
eigenfunctions $\psi$ so that  
\begin{equation}\label{expec M}
|\langle \TN(2,0)\psi,\psi \rangle |= \frac 1{2M^{1/2}} .
\end{equation}
\end{prop}
\begin{proof}
To construct $\psi$, we use the multiplicities in the spectrum: 
 From the formulas \eqref{eigenphases} for the eigenphases $\phi_{\eta,l}$ 
we see that $\phi_{\eta,l}=\phi_{\eta',l'} \bmod 1$ if and only if 
$(\eta')^2=\eta^2 \bmod D$,
and in addition 
$$
l'=l-\frac{\eta^2-(\eta')^2}D +b(\eta-\eta')\bmod M .
$$ 
In particular the multiplicity of $\phi_{\eta,l}$ is exactly
$$
\#\{\eta' \bmod D': (\eta')^2=\eta^2 \bmod D\}  ,
$$ 
which is independent of $l$. 
As a special case, we have $\phi_{1,-b} = \phi_{-1,b}$ if $D>2$.  

Now take 
\begin{equation}\label{eigenpsi}
\psi=\frac 1{\sqrt{2}}\big( \psi_{1,-b} + \psi_{-1,b} \big)  .
\end{equation}
Then since $\psi_{1,-b}$, $\psi_{-1,b}$ are orthonormal eigenfunctions
with the same eigenphase, $\psi$ is a normalized eigenfunction. 
We compute 
\begin{equation}\label{T(2,0)}
\begin{split}
\langle \TN(2,0)\psi,\psi \rangle =& \frac 12 \bigg( 
\langle \TN(2,0)\psi_{1,-b},\psi_{1,-b} \rangle +
\langle \TN(2,0)\psi_{1,-b},\psi_{-1,b} \rangle \\
&+
\langle \TN(2,0)\psi_{-1,b},\psi_{1,-b} \rangle +
\langle \TN(2,0)\psi_{-1,b},\psi_{-1,b} \rangle 
\bigg) .
\end{split}
\end{equation}

 By Lemma~\ref{lemT}, we have 
$\langle \TN(2,0)\psi_{\eta,l},\psi_{\eta',l'} \rangle = 0$ unless 
$\eta'+n_1 =\eta \bmod D$. Thus in our case if $D>2$ we see that all
but the second summand in \eqref{T(2,0)} are zero. As for that second 
summand, we see from Lemma~\ref{lemT} that in absolute value it equals 
$$
|\langle \TN(2,0)\psi_{1,-b},\psi_{-1,b} \rangle | = 
\frac{|G(-2b,-2b;M)|}{M} ,
$$
where the Gauss sum $G(-2b,-2b;M)$ is given by 
$$
G(-2b,-2b;M) =  \sum_{x\bmod M}e(\frac{-2b(x^2+x)}M) .
$$
In particular, if $M$ is odd then by Lemma~\ref{gauss sum} 
its absolute value is $\sqrt{M}$.  
Thus we find that if $M$ is odd then for our
eigenfunction $\psi$ in \eqref{eigenpsi} we have 
$$
|\langle \TN(2,0)\psi,\psi \rangle| = \frac 1{2M^{1/2}} ,
$$
as required.
\end{proof}

\subsection{Proof of Theorem \ref{thm lower}} 
We require the following construction: 
\begin{lem}\label{lem alpha}
Given an increasing positive function $g(x)$, 
there is an irrational $\alpha$, which has approximants 
$|\alpha-a/N|< 1/N$ with arbitrarily large $N$ so that 
$M=N/\gcd(a,N)$ is odd, and satisfying 
$$
M\gg g(N)^2 .
$$
\end{lem}
Once Lemma~\ref{lem alpha} is proved,  we will  then take $\psi$ as
in Proposition~\ref{special eigen}  and then since $M$ is odd, 
\eqref{expec M} holds so that  
$$
|\langle \TN(2,0)\psi,\psi \rangle |= \frac 1{2M^{1/2}} \gg \frac 1{g(N)} ,
$$
which will conclude the proof of Theorem~\ref{thm lower}. \qed

\subsection{Continued fractions} 
To prove Theorem~\ref{thm lower all} and Lemma~\ref{lem alpha}, 
we first review some basic facts about continued fractions; see
\cite{Hardy79} for details. 
Give a sequence of integers $a_0,a_1,a_2,\dots$ with $a_i\geq 1$ if
$i\geq 1$, consider the (finite) continued fraction
$$
[a_0;a_1,a_2,\dots,a_n]=
a_0+\frac 1{a_1+ \frac 1{\ddots+ a_n}} .
$$ 
%ALTERNATIVE
%$$
%[a_0;a_1,a_2,\dots,a_n]= a_0+ \cfrac1\\a_1+\cfrac 1\\ \ddots +\cfrac 1
%a_n %\endcfrac
%$$

The ``partial convergents'' $p_n$, $q_n$ are defined through the recursion
\begin{equation}\label{pnqn}
p_{n+1}=a_{n+1}p_n+p_{n-1},\qquad q_{n+1}=a_{n+1}q_n+q_{n-1},\quad
n\geq 1
\end{equation}
with initial conditions $p_0=a_0$, $q_0=1$, $p_1=a_1 a_0+1$, $q_1=a_1$. 
We have  
$$
\frac {p_n}{q_n} = [a_0;a_1,a_2,\dots,a_n] .
$$
The partial convergents satisfy the relation 
$$
p_n q_{n-1} - p_{n-1}q_n = (-1)^{n-1} ,
$$
from which it follows that $p_n$ and $q_n$ are  co-prime, and that 
$q_{n-1}$ and $q_n$ are co-prime. In particular at least one of
$q_{n-1}$, $q_n$ is {\em odd}. Another consequence is  
\begin{equation}\label{cauchy}
\frac{p_n}{q_n} - \frac{p_{n-1}}{q_{n-1}} = \frac{(-1)^{n-1}}{q_{n-1}q_n} .
\end{equation}

We now construct the continued fraction $\alpha:=[a_0;a_1,a_2,\dots]$
as the limit 
$$
\alpha:=[a_0;a_1,a_2,\dots]=a_0+\frac 1{a_1+ \frac 1{a_2+\ddots}} =
\lim_{n\to\infty} \frac {p_n}{q_n}
$$
(the limit exists by virtue of  \eqref{cauchy}). It defines an
irrational number.  

Conversely, for any irrational $\alpha$, set $\alpha_0=\alpha$, and
for $n\geq 0$ define  integers $a_n$ and reals $\alpha_{n+1}>1$ by 
$\alpha_n=a_n+ 1/\alpha_{n+1}$. The integers $a_n$'s are called the
``partial quotients'' of $\alpha$, and are positive if $n\geq 1$. 
Then 
$$
\alpha = [a_0;a_1,a_2,\dots,a_{n-1}+\frac 1{\alpha_n} ]=
a_0+\frac 1{a_1+ \ddots +\frac 1{ a_{n-1}+\frac 1{\alpha_n} } }  .
$$
We have 
$$
\alpha = \frac{\alpha_{n+1}p_n+p_{n-1}}{\alpha_{n+1}q_n+q_{n-1}}
$$
and 
$$
\alpha  - \frac {p_n}{q_n}=
\frac{(-1)^n}{q_n(\alpha_{n+1}q_n+q_{n-1})} .
$$
In particular, since $\alpha_{n+1}>a_{n+1}$ and $q_n\geq 1$  one gets 
\begin{equation}\label{approx} 
|\alpha  - \frac {p_n}{q_n}| < \frac 1{a_{n+1}q_n^2} .
\end{equation}

\subsection{Proof of Theorem \ref{thm lower all}} 
By Proposition~\ref{special eigen}, given $\alpha$ it suffices to find
arbitrarily large $N$   
and approximants $|\alpha-a/N|< 1/N$ so that $M=N/\gcd(a,N)$ is odd and
satisfies $N=M^2$. To do so, let $\frac bM=\frac{p_n}{q_n}$ be a partial
convergent with $M=q_n$ odd. Since at least one of $q_{n-1}$, $q_n$ is
odd, there are infinitely many such $M$. 
Set 
$$
N:=q_n^2=M^2, \quad a:=p_n q_n ,
$$ 
so that $D:=\gcd(a,N)=q_n$. We have 
$$
|\alpha-\frac aN|=|\alpha-\frac{p_n}{q_n}| < \frac 1{q_n^2}=\frac 1N ,
$$ 
so all our requirements are satisfied. 
This proves Theorem~\ref{thm lower all}. \qed

\subsection{Proof of Lemma \ref{lem alpha}} 
We begin with a construction of an
irrational: 
\begin{lem}\label{Liouville}
 Given any positive increasing function $F(x)$ 
there is an irrational $\alpha$ so that there are arbitrarily large
$q_n$ and approximants $p_n/q_n$ so that $F(q_n)\leq a_{n+1}q_n^2$ and 
$$
|\alpha -\frac {p_n}{q_n} |<\frac 1{F(q_n)}  .
$$
Moreover, we can require that $q_{n}$ are all odd. 
\end{lem}
\begin{proof}
We define $\alpha$ through its continued fraction expansion, that is
via the partial quotients $a_n$. 
Set $a_0=0$, and $a_1\geq 1$ to be  integer with $a_1\geq F(1)$. 
We define the partial quotients $a_i$ inductively: 
Given $a_0, a_1,\dots,a_n$, %which are all {\em even}, 
we get the partial convergents
$p_n$, $q_n$, and now choose $a_{n+1}$ to be an integer so that
$a_{n+1} \geq F(q_n)/q_n^2$.  
Set $\alpha=[a_0;a_1,a_2,\dots]$. Then from \eqref{approx} 
$$
|\alpha -\frac {p_n}{q_n} |<\frac 1{a_{n+1}q_n^2} \leq \frac 1{F(q_n)} ,
$$
by our choice of $a_{n+1}$. 
Since at least one of every pair of consecutive $q_n$'s is odd, we get
infinitely many $p_n/q_n$ satisfying our requirements. 
\end{proof}

To conclude the proof of Lemma~\ref{lem alpha}, that is 
to find the required $\alpha$, 
set $G=g^2$, which is increasing.
Then let $F=G^{-1}$ be the inverse function to $G$ 
which exists since $G$ is increasing, and is positive. 

Using Lemma~\ref{Liouville}, we construct an irrational $\alpha$ 
whose partial convergents satisfy 
$$
|\alpha -\frac {p_n}{q_n} |<\frac 1{a_{n+1}q_n^2} \leq \frac 1{F(q_n)} .
$$
Now take $n$ so that $q_n$ is odd (there are infinitely many such
$n$'s) and set $M:=q_n$, $b:=p_n$ (these are co-prime), 
and 
$$
N=a_{n+1}q_n^2, \quad a=a_{n+1}q_n p_n 
$$ 
so that $D:=\gcd(a,N)=a_{n+1}q_n$, and  $|\alpha- a/N|< 1/N$.  
 Finally, $M\leq G(N)=g(N)^2$ because  $F(q_n) \leq a_{n+1}q_n^2=N$ and since
$G$ is increasing, $M=q_n=G(F(q_n))\leq G(N) = g(N)^2$ as required. \qed

\newpage


\begin{thebibliography}{99}

\bibitem{CdV85}
Y.\,Colin de Verdi\`ere, Ergodicit\'e et 
fonctions propres du laplacien, {\em Comm. Math. Phys.} {\bf 102} (1985)
497-502.

\bibitem{Bouzouina96}
A.\,Bouzouina and S.\,De Bi\`evre,
Equipartition of the eigenfunctions of quantized
ergodic maps on the torus, 
{\em Comm. Math. Phys.} {\bf 178} (1996) 83-105. 

\bibitem{DeBievre96}
S.\,De Bi\`evre, M.\,Degli Esposti
and R.\,Giachetti, Quantization of a class of
piecewise affine transformations on the torus, 
{\em Comm. Math. Phys.} {\bf 176} (1996) 73-94.

\bibitem{Degli93} M. Degli Esposti, Quantization of the orientation
preserving automorphisms of the torus, {\em Ann. Inst. Poincar\'e} {\bf 58}
(1993) 323-341.
 
\bibitem{Degli95} M.\,Degli Esposti, S.\,Graffi and S.\,Isola, Classical
limit of the quantized hyperbolic toral automorphisms, {\em Comm. Math.
Phys.} {\bf 167} (1995) 471-507. 

\bibitem{Furstenberg61}
H.\,Furstenberg, Strict ergodicity and transformation of the torus, 
{\em Amer. J. Math.} {\bf 83} (1961) 573-601. 

\bibitem{Hannay80}
J.H.\,Hannay and M.V.\,Berry, 
Quantization of linear maps on a torus --- Fresnel
diffraction by a periodic grating,
{\em Physica} D {\bf 1} (1980) 267-290. 

\bibitem{Hardy79}
G.H.\,Hardy and E.M.\,Wright, 
{\em An introduction to the theory of numbers}
(The Clarendon Press, Oxford University Press, New York, 1979).

\bibitem{Helffer87}
B.\,Helffer, A.\,Martinez and D.\,Robert, 
Ergodicit\'e et limite semi-classique,
{\em Comm. Math. Phys.} {\bf 109} (1987) 313-326.

\bibitem{Jakobson94}
D.\,Jakobson, 
Quantum unique ergodicity for Eisenstein series on $\PSL_2(\ZZ)
\backslash\PSL_2(\RR)$, {\em Ann. Inst. Fourier} (Grenoble) {\bf 44} 
(1994) 1477-1504. 

\bibitem{Keating91}
J.P.\,Keating, 
The cat maps: quantum mechanics and classical motion. 
{\em Nonlinearity} {\bf 4} (1991) 309-341. 

\bibitem{Klimek97} S.\,Klimek, A.\,L\'esniewski, N.\,Maitra and R.\,Rubin, 
Ergodic properties of quantized toral automorphisms, 
{\em J. Math. Phys.} {\bf 38} (1997) 67-83.
 
\bibitem{Luo95}
W.\,Luo and P.\,Sarnak, 
Quantum ergodicity of eigenfunctions on $\PSL_2(\ZZ)\backslash 
\HH^2$, {\em Inst. Hautes \'Etudes Sci. Publ. Math.} {\bf 81} (1995)
207-237. 

\bibitem{vonNeumann29}
J.\,von Neumann, 
Beweis des Ergodensatzes und des $H$-Theorems in der neuen Mechanik,
{\em Zeitschr. f. Physi} {\bf 57} (1929) 30-70
({\em Collected Works} Vol.~1, Pergamon Press, Oxford 1961).

\bibitem{Rudnick94}
Z.\,Rudnick and P.\, Sarnak, The behaviour of eigenstates of arithmetic
hyperbolic manifolds, {\em Comm. Math. Phys.} {\bf 161} (1994) 195-213.

\bibitem{Schnirelman74}
A.I.\,Schnirelman, Ergodic properties of eigenfunctions,
{\em Uspehi Mat. Nauk} {\bf 29} (1974) 181-182.

\bibitem{Zelditch87}
S.\,Zelditch, Uniform distribution of eigenfunctions on compact
hyperbolic surfaces, {\em Duke Math. J.} {\bf 55} (1987) 919-941. 

\bibitem{Zelditch97}
S.\,Zelditch, Index and dynamics of quantized contact transformations,
{\em Ann. Inst. Fourier} (Grenoble) {\bf 47} (1997) 305-363. 
\end{thebibliography}
\end{document}